\documentclass{aa}
\usepackage{txfonts}
\usepackage{graphicx}

\title{Effects of alpha particles on the angular momentum loss from the Sun}
\author{Bo Li \and  Xing Li}
\offprints{Bo Li, \email{bbl@aber.ac.uk}}
\date{Received / Accepted}
\institute{Institute of Mathematical and Physical Sciences, University of Wales Aberystwyth,
   SY23 3BZ UK}
\titlerunning{Solar angular momentum loss}
\authorrunning{Bo Li \& Xing Li}
\begin{document}
\abstract
{}
{The classic Weber-Davis model  of the solar wind
    is reconsidered by incorporating alpha particles 
     and by allowing the solar wind to flow out of the equatorial plane
     in an axisymmetrical configuration.
}
{
In the ion momentum equations of the solar wind, the ion gyro-frequency
    is many orders of magnitude higher than any other
    frequency. 
This requires that the difference between proton and
    alpha velocity vectors be aligned with the background magnetic
    field. 
With the aid of this alignment condition, the governing
    equations of the multi-fluid solar wind are derived
    from the standard transport equations. 
The governing equations are numerically solved along a
    prescribed meridional magnetic field line
    located at colatitude $70^\circ$ at 1AU and
    a steady state fast solar wind solution is found.
}
{
A general analysis concludes, in agreement with the Weber-Davis model, that
    the magnetic field helps the coronal plasma
    to achieve an effective corotation out to the 
    Alfv\'enic radius, where the poloidal Alfv\'enic Mach number $M_T$ equals unity
    ($M_T$ is defined by equation~(\ref{eq:mach})).
The model computations show that, 
    magnetic stresses predominate the angular momentum loss of the Sun.
For the fast wind considered, the proton contribution to the angular momentum loss,
    which can be larger than the magnetic one,
    is almost completely canceled by the alpha particles
    that develop an azimuthal speed
    in the direction opposite to the solar rotation.
The Poynting flux associated with the azimuthal components
    is negligible in the energy budget.
However, the solar rotation can play some role in reducing the 
    relative speed between alpha particles and protons for low latitude 
    fast solar wind streams  in interplanetary space.
}
{}

\maketitle

\section{Introduction}

The solar angular momentum loss rate ${\cal L}$ consists of the 
     particle contribution ${\cal L}_P$ and that contained
     in magnetic stresses ${\cal L}_M$.
The comparison of measurements of these quantities with models,
     the Weber-Davis analysis (\cite{WD:67}) in particular,
     has yielded divergent results.
Missions before Helios measured a
     total angular momentum flux ${\cal L}$
     consistent with the Weber-Davis model
     (about $10^{30}$~dyne cm sr$^{-1}$), 
     but the measured azimuthal angle of the bulk flow
     was generally greater than
     $1^\circ$ at 1~AU (or equivalently 7~km~s$^{-1}$ for an average
     slow wind of 400~km~s$^{-1}$)
     (see Pizzo et al.~\cite{Pizzo.etal:83} and references therein).
Such a large azimuthal flow speed implies that 
       particles play a far more important role than magnetic stresses
     in reducing the angular momentum of the Sun.
However, in the Weber-Davis model, 3/4 of the angular momentum flux at 1~AU
     is due to magnetic stresses.
The Helios data show that ${\cal L}$ is $0.2-0.3 \times 10^{30}$ dyne cm sr$^{-1}$,
      in which ${\cal L}_M$ is about $0.15-0.2\times 10^{30}$ dyne cm sr$^{-1}$
     (Pizzo et al.~\cite{Pizzo.etal:83}).
Although the measured magnitude of ${\cal L}$ is smaller than that computed in
    the Weber-Davis model,
    the distribution of angular momentum flux between particles and magnetic stresses
    is largely compatible with their prediction.
An equally important finding concerns further 
    distribution of ${\cal L}_P$
    between two major ion species in the solar wind, namely,
    protons and alpha particles.
Alpha particles are found to carry an angular momentum
    flux in the direction of counter-rotation with the Sun.
This flux is substantial enough to
    offset the proton contribution which could be
    comparable to the magnetic one.
This finding cannot be addressed by the Weber-Davis model
     where the solar wind was treated as a bulk flow and
     only protons were considered.

Apart from being essential in the problem of solar angular momentum loss,
     the azimuthal ion motions may also provide a possible
     means to resolve a long  standing  observational puzzle,
     namely that
     alpha particles are observed to flow faster than
     protons in the fast solar wind.
The differential streaming in the fast wind
     could be as pronounced as 
     $150$~km~s$^{-1}$ at 0.3~AU before decreasing to about
     $40$~km~s$^{-1}$ at 1~AU
     (Marsch et al.~\cite{Marsch.etal:82}).
Such a behavior has yet to be understood.
One possible mechanism is the coupling between the azimuthal
     and meridional motions, facilitated
     by the strong magnetic field 
     (McKenzie et al.~\cite{McKenzie.etal:79}; 
     Hollweg \& Isenberg~\cite{HI:81}).
Although the Poynting flux may still be negligible
     (cf. Acuna \& Whang~(\cite{AcunaWhang:76}),
      Alexander \& {de La Torre}~(\cite{AT:95}),
      Hu et al.(\cite{Hu.etal:03b})),
     the proposed coupling
     is expected to limit, at least to a non-trivial extent, the ion 
     differential streaming.
As pointed out by Hollweg \& Isenberg~(\cite{HI:83}),
     one shortcoming of the formulation of McKenzie et al.~(\cite{McKenzie.etal:79})
     is 
     that protons are privileged over alphas:
     The azimuthal magnetic field is assumed to be
     determined entirely by the protons 
     whose azimuthal flow is neglected.
Hence the formulation cannot properly account for the
     azimuthal dynamics of protons or alphas.
In addition, the formulae are applicable only to the equatorial plane
     where the slow solar wind prevails at solar minimum.

The goal of this paper is to extend the Weber-Davis model by including
     alpha particles self-consistently.
This approach allows us to assess the individual contributions of
     ion flows and magnetic stresses to the angular momentum loss of the Sun.
The effect of the coupling between azimuthal and meridional motions
     in limiting the proton-alpha differential streaming
     will also be explored quantitatively. 
Given that the differential streaming is more prominent in the fast wind, 
     which in general flows out of the equatorial plane,
     it is necessary to
     formulate the model such that it
     treats both protons and alphas on an equal footing,
     and allows the solar wind to flow  outside the equatorial plane.
In this sense, this paper also extends
     the model of McKenzie et al.~(\cite{McKenzie.etal:79}).

The paper is organized as follows.
The derivation of the governing equations is given in the appendix.
Section~\ref{sec:model} details the physical model
    and also describes the
assumptions on the background poloidal magnetic field and 
    the ion heating mechanism.
A general analysis is then given in section~\ref{sec:analysis}.
Section~\ref{sec:numres} presents the numerical results
    and the effect of the solar rotation.
In section~\ref{sec:conc}, the main results are summarized.

\section{Model}\label{sec:model}

The solar wind is assumed to consist of electrons ($e$), protons ($p$) 
    and alpha particles ($\alpha$).
Since the role of alpha particles is not necessarily minor, their contribution
    has to be self-consistently taken into account (Li et al.\cite{Li.etal:97}).
This is done by rewriting the momentum equations
    (Schunk~\cite{Schunk:77}) in the flux tube frame, instead of
    the standard spherical coordinate system ($r, \theta, \phi$).
Central to the derivation is that the ion-cyclotron frequency 
    $\Omega_k = Z_k e B_l/m_k c$
    is many orders of magnitude higher than any other frequency present in
    the ion momentum equations  (McKenzie et al.\cite{McKenzie.etal:79}).
Here $m_k$ is the mass of ion species $k$ ($k=p, \alpha$),
     $Z_k$ is the charge of species $k$ in units of the electron charge $e$,
     $B_l$ denotes the meridional magnetic field strength, 
     and $c$  is the speed of light. 
The derivation is provided in the appendix, the resulting governing equations
     are quoted here to save space.

\subsection{Governing Equations}
\label{sec:goveqs}
  
As described in Appendix~A, in addition to the impicit assumptions in deriving 
    the five-moment transport equations
    (Schunk~\cite{Schunk:77}), we make the following assumptions:
\begin{enumerate}
\item Axial symmetry is assumed ($\partial/\partial\phi\equiv 0$);
\item The electron inertia is neglected in the electron momentum equation;
\item Quasi-neutrality is assumed;
\item Both viscosity and resistivity are neglected;
\item Quasi-zero current is assumed, the only exception to this occurs when ion
      momentum equations are derived;
\item  The electric field in the magnetic induction law is  
       convected by electrons, i.e., contributions like Hall
       effects are neglected;  
\item  The Spitzer law is used for the field-aligned electron heat flux,
          and the ion heat fluxes are neglected;
\item  We are only interested in steady state solutions.
       However, time-dependent equations are solved to yield steady state solutions.
\end{enumerate}
Given these assumptions, the governing equations take the form
\begin{eqnarray}
\frac{\partial n_k}{\partial t} 
    &+&\frac{1}{a}\frac{\partial}{\partial l}
         \left(n_k v_{k l} a\right) =0, \label{eq:nk}\\
\frac{\partial v_{k l}}{\partial t} 
    &+& v_{k l}\frac{\partial v_{k l}}{\partial l}    
       +\frac{1}{n_k m_k} \frac{\partial p_k}{\partial l} 
       +\frac{Z_k}{n_e m_k}\frac{\partial p_e}{\partial l}  \nonumber \\
    &+& \frac{G M_\odot}{r} \frac{\partial}{\partial l}\ln r 
       -\frac{1}{n_k m_k}\left(\frac{\delta M_{k l}}{\delta t}
             +\frac{Z_k n_k}{n_e}\frac{\delta M_{e l}}{\delta t}\right)
\nonumber \\
    &-&  v_{k\phi}^2\frac{\partial }{\partial l}\ln r\sin\theta  \nonumber\\
    &+&  \tan\Phi\left[ 
         v_{k l}\left(\frac{\partial}{\partial l} v_{k\phi}
        +v_{k\phi}\frac{\partial}{\partial l}\ln r\sin\theta \right)\right. \nonumber\\
    &-& \left. 
        \frac{1}{n_k m_k}\left(\frac{\delta M_{k\phi}}{\delta t}
             +\frac{Z_k n_k}{n_e}\frac{\delta M_{e\phi}}{\delta t}\right)\right] 
        =0, \label{eq:vks} \\
\frac{\partial T_e}{\partial t}
    &+&v_{e l}\frac{\partial T_e}{\partial l} 
       +\frac{(\gamma-1)T_e}{a}\frac{\partial}{\partial l}(v_{e l}a)
       -\frac{\gamma-1}{n_e k_B}\frac{\delta E_e}{\delta t} \nonumber \\
    &-&\frac{\gamma-1}{n_e k_B a}
      \frac{\partial}{\partial l}(a \kappa_e T_e^{5/2}
          \frac{\partial T_e}{\partial l}\cos^2\Phi)
     =0, \label{eq:te}\\
\frac{\partial T_k}{\partial t} 
    &+& v_{k l}\frac{\partial T_k}{\partial l} 
       +\frac{(\gamma-1)T_k}{a}\frac{\partial}{\partial l}(v_{k l}a)
       -\frac{\gamma-1}{n_k k_B}\frac{\delta E_k}{\delta t} \nonumber \\
    &-&\frac{\gamma-1}{n_k k_B} Q_k
       =0, \label{eq:tk}\\
\frac{\partial}{\partial t}v_{p \phi}
    &+&  v_{p l} \left(\frac{\partial}{\partial l} v_{p\phi}
       +v_{p\phi}\frac{\partial}{\partial l}\ln r\sin\theta\right) \nonumber\\
    &+& \frac{n_\alpha m_\alpha}{n_p m_p} v_{\alpha l}
        \left(\frac{\partial}{\partial l} v_{\alpha\phi}
            +v_{\alpha\phi}\frac{\partial}{\partial l}\ln r\sin\theta\right)  \nonumber\\
    &-&\frac{B_l}{4\pi n_p m_p}\left(\frac{\partial}{\partial l} B_\phi
       +B_\phi \frac{\partial}{\partial l}\ln r\sin\theta\right) =0, \label{eq:vphi}\\
\frac{\partial}{\partial t}B_{\phi}
    &+& \frac{r\sin\theta}{a}\frac{\partial}{\partial l}
         \left[\frac{a}{r\sin\theta}\left(B_\phi v_{e l}-v_{e\phi}B_l\right)\right]
       =0, \label{eq:bphi} \\
&& (v_{\alpha \phi}-v_{p\phi})=\frac{B_\phi}{B_l}(v_{\alpha l}-v_{p l}), \label{eq:flowdiff}
\end{eqnarray}
where $n_{\sf s}$, $\vec{v}_{\sf s}$ and $T_{\sf s}$ denote the number density,
     velocity and temperature of species ${\sf s}$ (${\sf s}=e,p,\alpha$), 
     respectively.
The species pressure is $p_{\sf s}= n_{\sf s} k_B T_{\sf s}$, 
     where $k_B$ is the Boltzmann constant.
By assuming quasi-neutrality 
    and quasi-zero current, we have $n_e=\sum_{k}Z_k n_k$ and
    $\vec{v}_e= \sum_{k}Z_k n_k \vec{v}_k/ n_e$ ($k = p, \alpha$). 
$G$ is the gravitational constant, $M_\odot$ is the solar mass,
    and $\gamma=5/3$ is the adiabatic index.
Coordinate $l$ measures the arclength of the poloidal magnetic field line
    from the footpoint at the coronal base.
Both the heliocentric distance $r$ and colatitude $\theta$
    are  evaluated along the poloidal magnetic field.
The cross-sectional area of the flux tube, $a$, scales as 
    $a \propto 1/B_l$.
$\Phi$ is the magnetic azimuthal angle, defined by 
    $\tan\Phi=B_\phi/B_l$.
The poloidal magnetic field $B_l$ and the heat deposition to ion species $k$, 
    denoted by $Q_k$, will be specified in
    sections~\ref{sec:bkmf} and \ref{sec:adhocht}, respectively.

The energy and momentum exchange rates $\delta E_{\sf s} /\delta t$
    and $\delta \vec {M}_{\sf s}/\delta t$ are due to
    Coulomb collisions of species ${\sf s}$ with all the remaining
    ones (Schunk~\cite{Schunk:77}), 
\begin{eqnarray}
\frac{\delta \vec{M}_{\sf s}}{\delta t} &=& 
    \sum_{\sf t} n_{\sf s} m_{\sf s}\nu_{\sf st}\Phi_{\sf st}
        \left(\vec{v}_{\sf t}-\vec{v}_{\sf s}\right),  
    \label{eq:deltaM}\\
\frac{\delta E_{\sf s}}{\delta t} &=& 
    \sum_{\sf t} \frac{n_{\sf s} m_{\sf s} \nu_{\sf st}}{m_{\sf s}+m_{\sf t}}
    \left[3k_B (T_{\sf t}-T_{\sf s})\Psi_{\sf st} 
          +m_{\sf t}(\vec{v}_{\sf t}-\vec{v}_{\sf s})^2\Phi_{\sf st}\right] .
    \label{eq:deltaE} 
\end{eqnarray}    
Expressions for the collision frequency $\nu_{\sf st}$ as well as
    correction factors $\Phi_{\sf st}$ and $\Psi_{\sf st}$
    have been given by, e.g., Li et al.~(\cite{Li.etal:97})
    and will not be repeated here. 
In the computation, the Coulomb logarithm $\ln\Lambda$ is taken to 
     be $21$.
The electron thermal conductivity $\kappa_e$ in Eq.~(\ref{eq:te}) is
    $7.8\times 10^{-7}$ {erg}~{K}$^{-7/2}$~{cm}$^{-1}$~s$^{-1}$
   (Spitzer~\cite{Spitz:62}).

\subsection{Background poloidal magnetic field}
\label{sec:bkmf}
  
To avoid complications associated with the cross-field
     force balance, 
     we choose to prescribe the background poloidal magnetic field
     by adopting an analytical model given in 
     Banaszkiewicz et al.~(\cite{Bana.etal:98}).
In the present implementation, the model magnetic field consists of
     dipole and current-sheet components only.
A set of parameters $M=3.6222$, $Q=0$, $K=1.0534$ and $a_1=2.5$ are
     chosen such that the last open magnetic field line is anchored
     at $\theta=40^\circ$ on the Sun, 
     and the poloidal magnetic field strength is 
     3.3$\gamma$ at $\theta=70^\circ$ at 1~AU, 
     compatible with
     Ulysses measurements 
     (Smith \& Balogh~\cite{SmithBalogh:95}).

\begin{figure*}
 \centering
 \includegraphics[width=1.\textwidth]{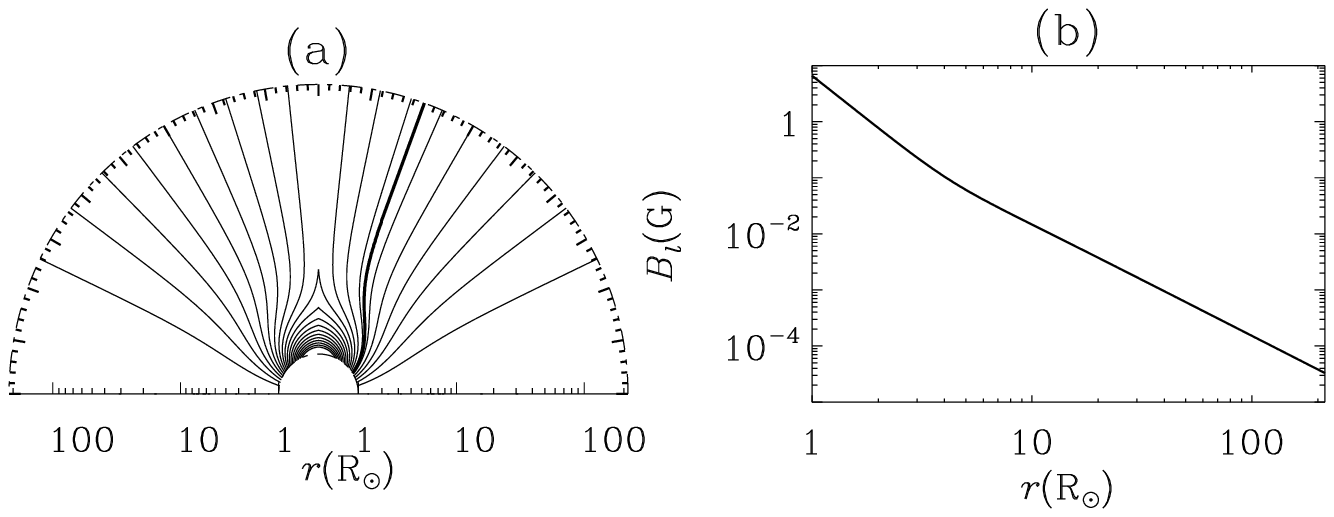}
 \vskip 0.5cm
 \caption{
   (a): The poloidal magnetic field configuration given as contours of the 
         magnetic flux function.
   The equator points upward.
   The line of force on which the model equations are solved is displayed by the
         thick contour.
   This field line is located at $\theta=70^\circ$ at 1~AU and originates from
         $31.5^\circ$ on the Sun.
   (b): Radial distribution of the poloidal magnetic field strength $B_l$
         along the designated field line.
   At 1~AU, $B_l$ is $3.3\gamma$.
  }
\label{fig:mf}
\end{figure*}

Figure~\ref{fig:mf}a shows the magnetic field configuration
    in the meridional plane.
The thick solid line represents the field line along which we will find
    solar wind solutions.
This field line is rooted at colatitude $31.5^\circ$ on the Sun,
    and reaches $70^\circ$ at 1~AU, which corresponds to the edge of
    the fast stream observed by Ulysses(McComas et al.~\cite{McComas.etal:00}).
Plotted in Fig.~\ref{fig:mf}b is the radial profile of
    the poloidal magnetic field strength $B_l$ along
    the designated field line.

\subsection{Ion heating}
\label{sec:adhocht}

To produce fast solar wind solutions,
     an empirical energy flux, launched from the Sun and in the direction of $\vec{B}$,
     is assumed to heat ions only.
This energy flux is assumed to dissipate
     at a rate $Q$ with a characteristic length~$l_d$, i.e.,  
\begin{equation}
 Q=F_{E} \frac{B_l}{B_{l E} l_d}\exp\left(-\frac{l}{l_d}\right), 
\label{eq:Q}
\end{equation}
    where $F_E$ is the input empirical flux scaled to
    the orbit of the Earth, $R_E = 215$~$R_\odot$,
    $R_\odot$ being the solar radius.
Moreover, $B_{l E}$ is the poloidal magnetic field strength at $R_E$.
$Q$ is then assumed to be apportioned between protons
    and alpha particles by
\begin{eqnarray}
 Q_{\alpha}= \frac{\Delta}{1+\Delta}, Q_p= \frac{1}{1+\Delta}, %\nonumber\\
 \Delta=\frac{\rho_{\alpha}}{\rho_p} \chi , \label{eq:distQ}
\end{eqnarray}    
    where $\rho_k = n_k m_k$ ($k=p, \alpha$) is the ion mass density,
    and $\chi$ is 
    a parameter indicating the degree by which the alpha particles
    are preferentially heated,
    with $\chi\equiv 1$ standing for the neutral heating:
    heating rate per ion is proportional to its mass. 

In the computations, the following parameters 
\begin{eqnarray*}
F_{E} = \mbox{1.8~erg~cm$^{-2}$~s$^{-1}$}, \hskip 0.2cm
 l_d = \mbox{1.35~}R_\odot , \\
\chi=\frac{\chi_{c}+0.8}{2}-\frac{\chi_{c}-0.8}{2} 
   \tanh\left(\frac{r-5R_\odot}{0.3 R_\odot}\right), \hskip 0.2cm \chi_c=1.5 
\end{eqnarray*}
    are chosen to yield a fast solar wind solution. 
As can be seen, $\chi$ varies smoothly from $\chi_c$
   in the inner corona to $0.8$ far from the Sun with
   a rather steep transition occurring at 5~$R_\odot$.
A  preferential heating that favors alpha particles in the inner
   corona ($\chi_c>1$) is necessary to produce a positive
   relative speed $v_{\alpha l}-v_{p l}$.

\section{General Analysis}
\label{sec:analysis}

Before solving equations (\ref{eq:nk}) to (\ref{eq:flowdiff}) to find solar wind solutions,
   one can conduct an analysis to reach some general conclusions.

\subsection{Alignment conditions}
\label{sec:align}
Equation~(\ref{eq:bphi}) derives from the $\phi$ component of the magnetic induction law.
For a steady state, it can be integrated to yield 
\begin{eqnarray}
v_{e\phi} 
    -\Omega r\sin\theta=\frac{B_\phi}{B_l} v_{e l}. \label{eq:aligne}
\end{eqnarray}    
The constant of integration $\Omega$ can be
    identified as the angular rotation rate of the flux tube.
%Once the differential rotation is taken into account, $\Omega$ should be different for
%    flux tubes anchored at different latitudes on the Sun.
Combining Eq.~(\ref{eq:aligne}) and ~(\ref{eq:flowdiff}) , one finds
\begin{eqnarray}
v_{p\phi} 
    -\Omega r\sin\theta=\frac{B_\phi}{B_l} v_{p l} , \hspace{1cm}
v_{\alpha\phi}
    -\Omega r\sin\theta=\frac{B_\phi}{B_l} v_{\alpha l} . \label{eq:align}
\end{eqnarray}
That is, in the frame strictly corotating with the Sun, 
    all species (electrons, protons and alpha particles)
    flow along the magnetic field.
The alignment conditions were first recognized by 
    Parker~(\cite{Parker:58}), and have been extended to
    general axisymmetrical MHD flows by, e.g., 
    Low \& Tsinganos~(\cite{LowTsing:86}) and Hu et al.(\cite{Hu.etal:03b}).

\subsection{Angular momentum conservation law}
\label{sec:angmom}
In a steady state, equation~(\ref{eq:vphi}) leads to 
\begin{eqnarray}
r\sin\theta\left[v_{p\phi}+\frac{\rho_\alpha v_{\alpha l}}{\rho_p v_{p l}} v_{\alpha\phi}
    -\frac{B_\phi B_l}{4\pi\rho_p v_{p l}}
  \right] = L, \label{eq:angcons}
\end{eqnarray}
    where the tube invariant $L$ comes from the integration.
The physical meaning of $L$ can be better seen by noting that the constant
\begin{eqnarray}
{\cal L} &=& \rho_p v_{p l} L \frac{a}{a_E} R_E^2 =  {\cal L}_p + {\cal L}_\alpha +{\cal L}_M \label{eq:capL}
\end{eqnarray}
   is the angular momentum loss per solid angle, where
\begin{eqnarray}
{\cal L}_k = r\sin\theta \rho_k v_{k l} v_{k\phi} \frac{a}{a_E}R_E^2,  \hskip 0.2cm
{\cal L}_M =  -r\sin\theta\frac{B_\phi B_l}{4\pi} \frac{a}{a_E}R_E^2, \label{eq:Lspec}
\end{eqnarray}
    with $k=p, \alpha$.
Subscript $E$ denotes values evaluated at $R_E=1$~AU.
Obviously, both outflowing particles and  magnetic stresses
    contribute to the angular momentum flux.

The conservation law for the angular momentum, Eq.~(\ref{eq:angcons}),
    is valid for an arbitrary flux tube in an azimuthally symmetric
    solar wind.
The single-fluid version (or equivalently the two-fluid one) of this conservation
    law has already been obtained by, e.g., 
    Low \& Tsinganos~(\cite{LowTsing:86}) and Hu et al.(\cite{Hu.etal:03b}).

\subsection{Expressions for $v_{p \phi}$, $v_{\alpha \phi}$ and $B_\phi$}
\label{sec:vphibphi}
The alignment condition, Eq.~(\ref{eq:align}), 
    together with Eq.~(\ref{eq:angcons}) leads to
\begin{eqnarray}
v_{p\phi} &=& \frac{\Omega r\sin\theta}{M_T^2-1}
   \left[M_p^2\frac{L}{\Omega r^2\sin^2\theta}-1
    + M_\alpha^2\frac{v_{\alpha l}-v_{p l}}{v_{\alpha l}}\right], \label{eq:vpphi}\\
v_{\alpha\phi} &=& \frac{\Omega r\sin\theta}{M_T^2-1}
   \left[M_p^2\frac{L}{\Omega r^2\sin^2\theta}-1 \right. \nonumber\\
    &+&\left. M_p^2\frac{v_{\alpha l}-v_{p l}}{v_{p l}}
    \left(\frac{L}{\Omega r^2\sin^2\theta}-1\right)\right], \label{eq:viphi}\\
B_{\phi} &=& \frac{4\pi\rho_p v_{p l}}{B_l r\sin\theta}
   \frac{L-\left(1+\frac{\rho_\alpha v_{\alpha l}}{\rho_p v_{p l}}\right)
   \Omega r^2\sin^2\theta}{M_T^2-1}, \label{eq:bphia}
\end{eqnarray}
where $M_T$, $M_p$ and $M_\alpha$ are defined as
\begin{eqnarray}
M_T^2  = M_p^2+M_\alpha^2, \hskip 0.2cm 
 M_k^2=\frac{v_{k l}^2}{B_l^2/4\pi\rho_k}, 
\label{eq:mach}
\end{eqnarray}
with $k=p,\alpha$.  
The poloidal Alfv\'enic Mach number $M_T$ is thus comprised of
    both $M_p$ and $M_\alpha$. 

For the solar wind, $M_T \ll 1$ is valid near 1~$R_\odot$,
    but $M_T \gg 1$ holds at 1~AU.
Hence, there must exist a point between 1~$R_\odot$ and 1~AU where $M_T=1$.
At this location, which will be termed the Alfv\'enic point,
    $B_\phi$ is singular unless the numerator in Eq.~(\ref{eq:bphia}) vanishes,
\begin{eqnarray}
L = \left(1+\frac{\rho_\alpha v_{\alpha l}}{\rho_p v_{p l}}\right)
   \Omega r_a^2\sin^2\theta_a, \label{eq:L}
\end{eqnarray}
   where subscript $a$ denotes values at the Alfv\'enic point.
We have employed the fact that the ion mass flux ratio
   $\rho_\alpha  v_{\alpha l}/\rho_p v_{p l}$ is a constant.
The angular momentum loss per solid angle then becomes
\begin{eqnarray}
{\cal L} = \dot{M} \Omega r_a^2 \sin^2\theta_a, 
\end{eqnarray}
   where
\begin{eqnarray*}
\dot{M} = (\rho_p v_{p l}+\rho_\alpha v_{\alpha l})\frac{a}{a_E}R_E^2 
\end{eqnarray*}
   is the mass loss rate per solid angle of the solar wind.

Hence the conclusion of
    Weber \& Davis~(\cite{WD:67}) still holds: the magnetic field 
    helps the coronal plasma to achieve an effective corotation 
    to the Alfv\'enic point,
    as long as the poloidal Alfv\'enic Mach number $M_T$ is defined by Eq.~(\ref{eq:mach}).

\subsection{Energy conservation law}
\label{sec:enercons}

Combining the governing equations in the steady state, one can
   derive an energy conservation law,
\begin{eqnarray}
&& \frac{a}{a_E}\left[\rho_p v_{p l}\frac{v_{p l}^2+v_{p\phi}^2}{2}
   +\rho_\alpha v_{\alpha l}\frac{v_{\alpha l}^2+v_{\alpha\phi}^2}{2}
   \right] \nonumber \\
&+&  \left(\rho_p v_{p l}+\rho_\alpha v_{\alpha l}\right)\frac{a}{a_E}
     {G M_\odot}\left(\frac{1}{R_\odot}-\frac{1}{r}\right)\nonumber \\
&+& \frac{a}{a_E}
    \frac{B_\phi}{4\pi}\left(v_{e l}B_\phi-v_{e\phi}B_l\right) \nonumber \\
&+&  \frac{a}{a_E}\frac{\gamma}{\gamma-1}
    \left(p_e v_{e l} +  p_p v_{p l} +p_\alpha v_{\alpha l}\right) \nonumber\\
&-& \frac{a}{a_E}\kappa_e T_e^{5/2}\frac{\partial T_e}{\partial l}\cos^2\Phi \nonumber \\	
&+& \int_{r}^{R_E}\frac{a}{a_E} (Q_p+Q_\alpha) {\rm d} l = {\cal F}, \label{eq:enercons}
\end{eqnarray}
    where the constant 
    ${\cal F}$ is the total energy flux scaled to $R_E$.
The terms on the left hand side of Eq.~(\ref{eq:enercons}) correspond,
    respectively, to the kinetic and potential energy fluxes, 
    the Poynting flux, the enthalpy flux, the electron conductive flux,
    and the source term due to the heat deposition.
The ratio of the Poynting flux to ${\cal F}$ will be used to
    assess the
    relative importance of the Poynting flux in the energy budget.

\section{Numerical Results}
\label{sec:numres}

Equations (\ref{eq:nk}) to (\ref{eq:flowdiff}) 
    are solved by using a fully implicit numerical scheme
    (Hu et al.~\cite{Hu.etal:97}).
From an arbitrary initial guess, the equations are advanced 
    in time until a steady state is achieved.
The computational domain extends from 1~$R_\odot$ to 1.2~AU. 
At 1~$R_\odot$, ion densities as well as species temperatures are fixed, 
\begin{eqnarray*}
&& n_{p}=1.5\times 10^8 \mbox{cm}^{-3}, \hskip 0.2cm 
  (n_\alpha/n_p) = 0.06, \\\
&& T_{e}=T_{p} = T_{\alpha} = 10^6\mbox{K}, \hskip 0.2cm 
\end{eqnarray*}
    while $v_{p l}$ and $v_{\alpha l}$ are specified to ensure
    mass conservation.
$v_{p\phi}$ and $B_\phi$ are evaluated in accordance with
    equations~(\ref{eq:vpphi}) and (\ref{eq:bphia}), where $L$
    is computed at the grid point immediately adjacent to the base.
At the outer boundary ($1.2$~AU), all dependent variables
    are linearly extrapolated for simplicity.
We also take $\Omega=2.865\times 10^{-6}$~rad s$^{-1}$.
For the steady state solutions presented in this paper,
    the maximum relative errors in the conserved quantities
    are smaller than $1\%$.

\begin{figure*}
 \centering
 \includegraphics[width=1.\textwidth]{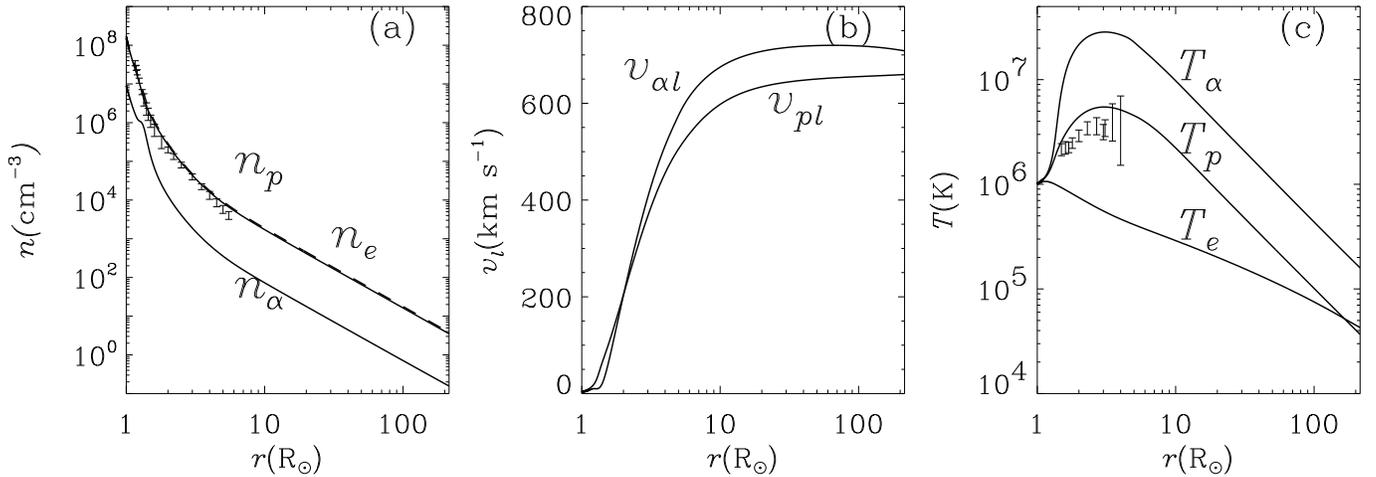}
 \vskip 0.5cm
 \caption{ Results derived from a 1.5-D 3-fluid solar wind model which incorporates
      the azimuthal components self-consistently.
      The radial distribution of 
    (a) the densities of protons $n_p$ and alpha particles $n_\alpha$ (solid lines),
        as well as electrons $n_e$ (dashed line),
    (b) the poloidal flow speeds of protons ($v_{pl}$) and alphas ($v_{\alpha l}$),
    and (c) the temperatures of electrons ($T_e$), protons ($T_p$)
        and alpha particles ($T_\alpha$).
    The error bars in (a) are the upper and lower limits for the electron
        density derived by Fisher \& Guhathakurta~(\cite{FG:95}).
    The error bars in (c) represent the uncertainties of UVCS measurements for the effective 
        proton temperature reported by Kohl et al.~(\cite{Kohl:98}).
    Please note that both measurements are made for polar coronal holes.
  }
\label{fig:refmodel}
\end{figure*}

Figure~\ref{fig:refmodel} displays the radial distribution of 
    (a) the species densities $n_e$, $n_p$ and $n_\alpha$, 
    (b) poloidal flow speeds $v_{p l}$ and $v_{\alpha l}$,
    and (c) species temperatures $T_e$, $T_p$ and $T_\alpha$.
%The error bars in Fig.\ref{fig:refmodel}a are the limits for the electron density 
%    derived from white light observations by
%    Fisher \& Guhathakurta~(\cite{FG:95}).
%Also plotted in Fig.\ref{fig:refmodel}c are the UVCS measurements for the effective 
%    proton temperature reported by
%    Kohl et al.~(\cite{Kohl:98}). 
%Note that both measurements are made for polar coronal holes.
The model yields the following parameters at 1~AU, 
\begin{eqnarray*}
&& n_p v_{p l} = 2.3 \times 10^8 \mbox{cm}^{-2} \mbox{s}^{-1}, \hskip 0.2cm
v_{p l} = 660\mbox{~km~s$^{-1}$},   \\
&& v_{\alpha l}-v_{p l} = 49\mbox{~km~s$^{-1}$},\hskip 0.2cm
n_\alpha/n_p = 0.0445 
\end{eqnarray*}
    which agree  well with the Ulysses observations 
    of the fast wind
    (McComas et al.~\cite{McComas.etal:00}).
In addition, the modeled electron density fits observations 
    reasonably well in the inner corona (Fig.~\ref{fig:refmodel}a).
However, without considering the non-thermal contribution, the modeled
    proton temperature is higher than that inferred from
    UVCS measurements
    (Fig.~\ref{fig:refmodel}c).
Moreover, $T_p$ and $T_\alpha$ at 1~AU are much smaller than
    values given by in situ measurements 
    (McComas et al.~\cite{McComas.etal:00}).
The poor match is due to the oversimplified heating function.
As for the poloidal flow speeds, the alphas initially fall behind the protons
    below 2~$R_\odot$ beyond which a positive
    $\Delta v_l =v_{\alpha l}-v_{p l}$ develops. 
$v_{\alpha l}$ reaches a maximum around 66~$R_\odot$,
    and starts to decrease thereafter.

\begin{figure*}
\centering 
 \includegraphics[width=.8\textwidth]{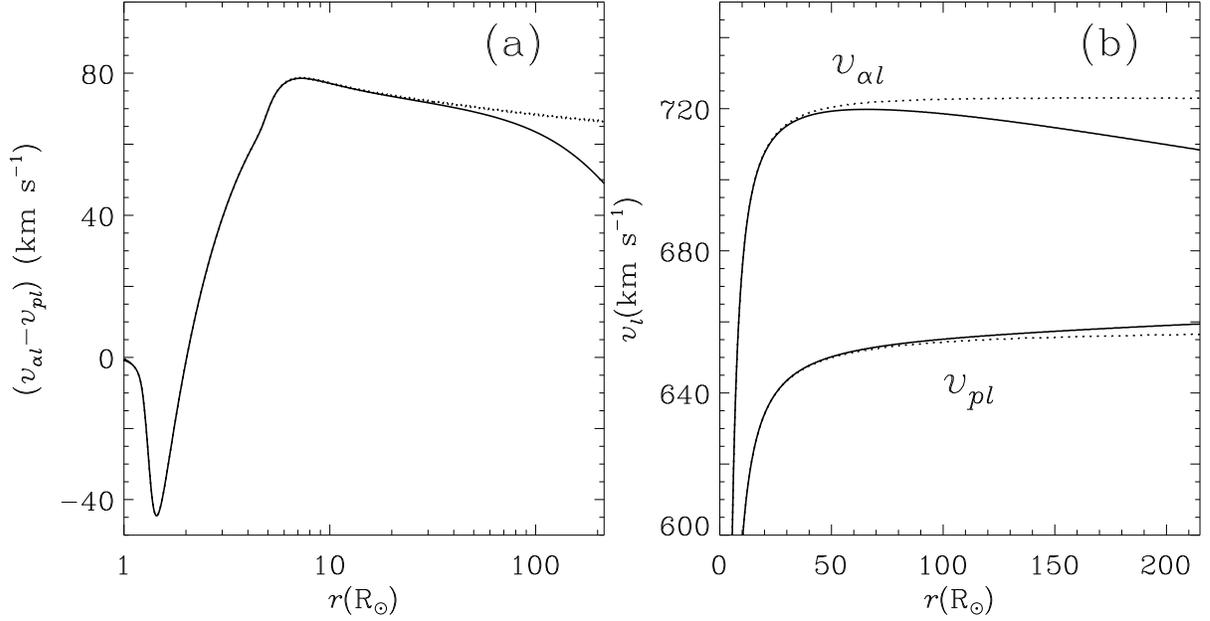}
 \vskip 0.5cm
 \caption{ 
Radial distributions of
    (a) the differential streaming, $v_{\alpha l}-v_{p l}$,
    and 
    (b) the poloidal flow speeds of protons ($v_{p l}$) and alpha particles ($v_{\alpha l}$).
Solid lines are used to plot the 1.5-D model, whereas dotted lines are
    used for the corresponding 1-D model which neglects the solar rotation.
  }
\label{fig:flowdiff}
\end{figure*}

To examine the differential streaming further,
    $\Delta v_l = v_{\alpha l}-v_{p l}$ is plotted in 
    Figure~\ref{fig:flowdiff}a.
The poloidal flow speeds of protons ($v_{p l}$) 
    and alpha particles ($v_{\alpha l}$) 
    are replotted in Fig.~\ref{fig:flowdiff}b
    (a different scale is used, see Fig.~\ref{fig:refmodel}b).
In addition, model results from the corresponding computation that
    neglects the solar rotation (i.e., $\Omega\equiv 0$)
    are plotted as dotted lines for comparison.
For the ease of description, we shall call the model with (without) azimuthal
    components the 1.5-D (1-D) model.
It is found that the effect of the azimuthal
    components on the poloidal dynamics can be adequately 
    represented by the flow speed profiles.
Below the local maximum of $78.6$~km~s$^{-1}$ at $7.28$~$R_\odot$,
    Fig.~\ref{fig:flowdiff}a shows no difference in the
    $\Delta v_l$ profile
    between 1-D and 1.5-D models.
The differential streaming, $\Delta v_l$, for both models plummets from nearly zero
    at the coronal base to 
    about $-44.6$~km~s$^{-1}$ at 1.44~$R_\odot$, 
    and rises thereafter to the local maximum.
Interestingly, in the 1-D model, beyond the local maximum 
    $\Delta v_l$ undergoes only a modest
    decrease to $66.3$~km~s$^{-1}$ at 1~AU,
    while in the 1.5-D model $\Delta v_l$ is $48.7$~km~s$^{-1}$ at 1~AU.
This further reduction in the differential streaming is achieved
    through a slight rise in the $v_{p l}$ profile 
    accompanied by a modest deceleration of alpha particles
    (Fig.~\ref{fig:flowdiff}b).

\begin{figure*}
 \centering
 \includegraphics[width=1.\textwidth]{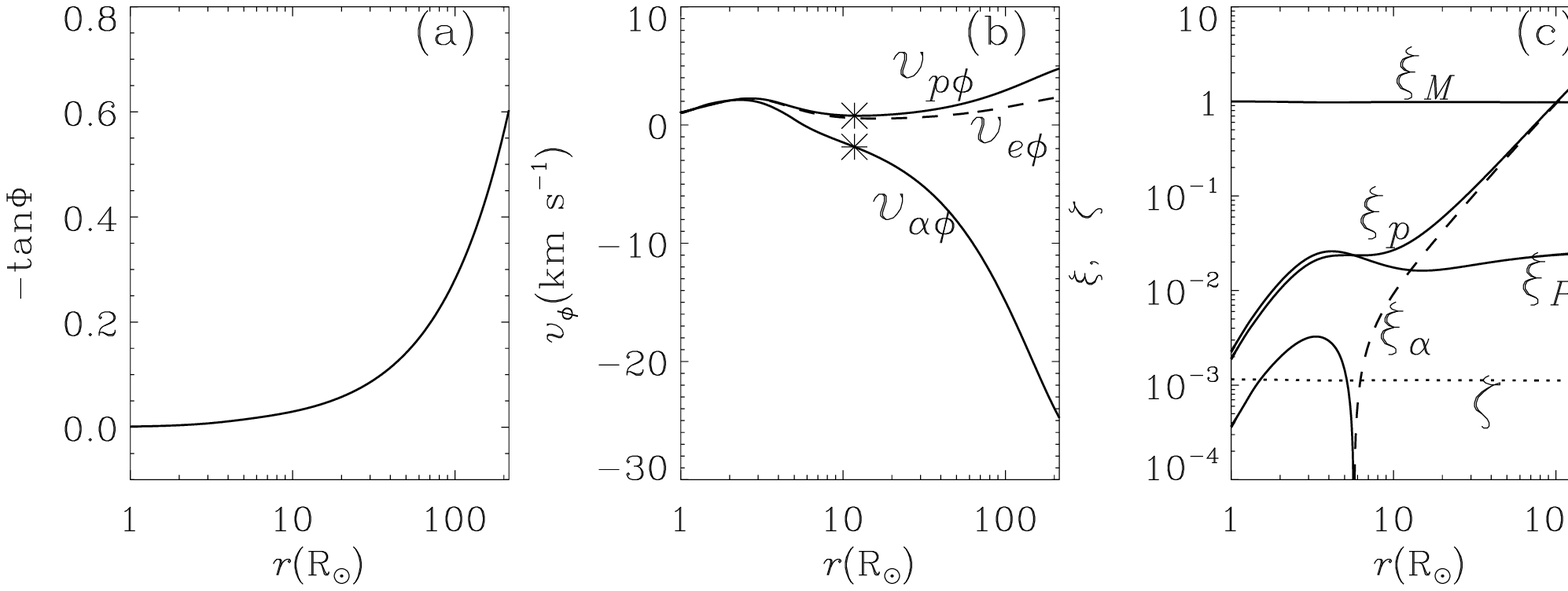}
 \vskip 0.5cm
 \caption{ Radial distributions of
    (a) $-\tan\Phi=-B_\phi/B_l$ where $\Phi$ is the magnetic azimuthal angle,
    (b) the azimuthal speeds of protons $v_{p\phi}$, alpha particles $v_{\alpha\phi}$ as well
        as electrons $v_{e\phi}$,
    (c) the relative importance of 
        the proton fluid $\xi_p = {\cal L}_p/{\cal L}$, 
        the alpha fluid $\xi_\alpha = {\cal L}_\alpha/{\cal L}$, 
        the sum of the two $\xi_P = \xi_p+\xi_\alpha$,
        and the magnetic stresses $\xi_M = {\cal L}_M/{\cal L}$
        in the total angular momentum loss of the Sun
        (please see Eq.~(\ref{eq:Lspec}) in text).
        In addition, the ratio of the Poynting flux to the total energy flux,
        $\zeta$, is plotted as dotted line. 
        The dashed line represents negative values.
    In panel (b), the asterisks denote the Alfv\'enic point, where the 
        poloidal Alfv\'enic Mach number (defined by Eq.~(\ref{eq:mach})) equals unity.
}
\label{fig:angmom}
\end{figure*}

This behavior is not surprising since  
   in the poloidal momentum equation (Eq.~(\ref{eq:vks})),
\begin{eqnarray}
&& v_{k l} \frac{\partial v_{k l}}{\partial l} 
       -v_{k\phi}^2 \frac{\partial}{\partial l}\ln r\sin\theta \nonumber\\
&+&\tan\Phi v_{k l}\left(\frac{\partial v_{k\phi}}{\partial l} 
       +v_{k\phi}\frac{\partial}{\partial l} \ln r\sin\theta\right) \nonumber\\
&=& \frac{\partial}{\partial l}\left(\frac{v_{k l}^2}{2} \sec^2\Phi\right)
  -\frac{\partial}{\partial l}\frac{\Omega^2 r^2\sin^2\theta}{2} \label{eq:corot}
\end{eqnarray} 
    can be obtained when the alignment condition,
    Eq.~(\ref{eq:align}), is used.
When viewed in the frame corotating with the Sun
    ($v_{k l} \sec\Phi$ is the ion speed seen in that frame), 
    the solar rotation ensures that all particles
    move in the same centrifugal potential ($\Omega^2 r^2\sin^2\theta/2$). 
Neglecting all other contributions, and taking the difference of the proton 
   and alpha version of Eq.~(\ref{eq:corot}), one arrives at 
\begin{eqnarray*}
\frac{\partial}{\partial l}\left[\left(v_{\alpha l}^2
     -v_{p l}^2\right)\sec^2\Phi\right] = 0 ,
\end{eqnarray*}  
or 
\begin{eqnarray}
(v_{\alpha l}^2-v_{p l}^2) \propto \cos^2\Phi.
\end{eqnarray}
With the development of the magnetic azimuthal angle, $\cos^2\Phi$ decreases
   monotonically with increasing distance (cf. Fig.~\ref{fig:angmom}a).
As a consequence, the differential streaming $\Delta v_l$ 
   decreases.
%The equivalence of developing governing equations in the inertial frame
%   and the corotating frame has been pointed out by
%   Hollweg \& Isenberg~(\cite{HI:81}) 
%   (see also, McKenzie \& Axford~\cite{MA:83}
%    and Hollweg \& Isenberg~\cite{HI:83}),
%   who also predicted in a slightly different manner
%   the effect of the solar rotation in limiting
%   the ion differential streaming.
Figure~\ref{fig:flowdiff} can be seen as a direct illustration of
    the effect of solar rotation in limiting the
    ion differential streaming, predicted by McKenzie et al.(\cite{McKenzie.etal:79})
    and Hollweg \& Isenberg~(\cite{HI:81}).

Figure~\ref{fig:angmom} displays the radial profiles of 
   (a) $-\tan\Phi=-B_\phi/B_l$,
   (b) the azimuthal speeds of protons ($v_{p\phi}$), alpha particles ($v_{\alpha\phi}$)
       and electrons ($v_{e\phi}$),
   and (c) the specific contribution of protons ($\xi_p = {\cal L}_p/{\cal L}$),
       alpha particles ($\xi_\alpha={\cal L}_\alpha/{\cal L}$)
       and the magnetic stresses ($\xi_M={\cal L}_M/{\cal L}$)
       to the angular momentum flux
   (cf. Eq.~(\ref{eq:capL})).
In addition, the sum $\xi_p +\xi_\alpha$, which gives the overall particle
   contribution $\xi_P$, is also plotted. 
Given in dotted line is $\zeta$, the ratio of the Poynting flux to the total energy flux
   (cf. Eq.~(\ref{eq:enercons})).
In Fig.~\ref{fig:angmom}c, the dashed line is used to  plot negative values.
The asterisks in Fig.~\ref{fig:angmom}b denote the 
    Alfv\'enic point, which is located at $r_a=11.8 R_\odot$.

From Fig.~\ref{fig:angmom}a, it is obvious that only beyond, say 10~$R_\odot$,
     does a spiral angle $\Phi$ develop.
This can be explained in view of equations~(\ref{eq:aligne}), (\ref{eq:align}):
     within 10~$R_\odot$ the left hand side is much smaller than the
     poloidal flow speed on the right hand side for any species.
On the other hand, in interplanetary space, 
   the species azimuthal speed
   is much smaller than $\Omega r\sin\theta$, 
   the Parker theory for the spiral magnetic field is recovered,
   i.e.,
   $\tan\Phi =B_\phi/B_l \approx -\Omega r\sin\theta/v_l$, where $v_l$ can be 
   taken as the poloidal speed of any species.

In the inner corona, both protons and alpha particles
    tend to corotate with the Sun:  $v_{p\phi}$ and 
    $v_{\alpha\phi}$ are positive (Fig.~\ref{fig:angmom}b).
The azimuthal speed of the alpha particles is slightly
    larger than that of the protons
    below 2~$R_\odot$, 
    and from there on, the alpha particles
    are gradually turned opposite to the solar rotation.
$v_{\alpha\phi}$ becomes negative beyond $5.71$~$R_\odot$,
    eventually $v_{\alpha\phi}$ reaches $-24.7$~km~s$^{-1}$ at 1~AU.
On the other hand, the proton azimuthal speed $v_{p\phi}$ increases from
    a local minimum of $0.78$~km~s$^{-1}$ at 12.5~$R_\odot$
    monotonically to $4.8$~km~s$^{-1}$ at 1~AU.

The behavior of the azimuthal flow speeds can be explained
    by Eqs.~(\ref{eq:vpphi}) and (\ref{eq:viphi}).
Near the coronal base,  both $M_p^2$ and $M_\alpha^2$ are far from unity.
It then follows from Eqs.~(\ref{eq:vpphi}) and (\ref{eq:viphi}) that
\begin{eqnarray}
v_{p\phi}\approx \Omega r\sin\theta, \hskip 0.2cm
  v_{\alpha\phi}\approx \Omega r\sin\theta  . \label{eq:corot2}
\end{eqnarray}
At $r \le 2\mbox{~}R_\odot$, a negative $\Delta v_l$
   makes $v_{\alpha\phi}$ slightly
   larger than $v_{p\phi}$.
For $r > r_a$, the solar wind expands almost radially.
As a result, $M_k^2/v_{k l} = (4\pi \rho_k v_{k l}/B_l)/B_l
     \propto r^2$ ($k=p,\alpha$) 
     holds fairly accurately.
The variation of $v_{pl}$ beyond $r_a$ is very modest.
We therefore have $M_p^2\gg 1$ for $r\gg r_a$.
From the identity
    $M_\alpha^2/M_p^2=\eta v_{\alpha l}/v_{p l} $, it
    follows that
    $M_\alpha^2$ is a substantial fraction of $M_p^2$
    given that
    the ion mass flux ratio $\eta=\rho_\alpha v_{\alpha l}/\rho_p v_{p l}$
     is 0.19 in this solution.
Hence close to 1~AU, the azimuthal speeds
    of both protons and alphas are 
    determined by the terms associated with the differential streaming
    in Eqs.~(\ref{eq:vpphi}) and (\ref{eq:viphi}), namely, for $r\gg r_a$,
\begin{eqnarray}
v_{p\phi}\approx \Omega r\sin\theta 
    \frac{\eta}{1+\eta {v_{\alpha l}/v_{p l}}}\frac{v_{\alpha l}-v_{p l}}{v_{p l}}, \nonumber \\
v_{\alpha\phi}\approx -\Omega r\sin\theta 
       \frac{1}{1+\eta {v_{\alpha l}/v_{p l}}}\frac{v_{\alpha l}-v_{p l}}{v_{p l}}. 
\end{eqnarray}
As a result, $v_{p\phi}/v_{\alpha\phi}\approx -\eta$ holds.
However, this asymptotic behavior of the ion azimuthal
     speeds for $r\gg r_a$ does not hold in general.
If the alpha abundance is far from unity, the azimuthal magnetic field 
     will be solely determined by protons, and $v_{p\phi}$ should behave like
     $v_{p\phi}\propto r^{-1}$ for $r\gg r_a$ when the differential streaming
     term in Eq.~(\ref{eq:vpphi}) is neglected.

Now let us move on to Fig.~\ref{fig:angmom}c.
It can be seen that, from the coronal base to 1~AU, 
    magnetic stresses play a dominant role
    in the total angular momentum budget, the particle
    contribution $\xi_P$ is no more than $2.6\%$.
However, the individual angular momentum flux carried 
   by protons or alpha particles is not necessarily small in magnitude.
As a matter of fact, protons contribute
    more to the total angular momentum flux
    than magnetic stresses do
    beyond $101$~$R_\odot$.
However the proton contribution
    is virtually canceled by the alpha particles that 
    counter-rotate with the Sun.
This can be understood in light of equations~(\ref{eq:vpphi}) and 
   (\ref{eq:viphi}).
As has been described, far away from the Alfv\'enic point, $r \gg r_a$,
    both $v_{p\phi}$ and $v_{\alpha\phi}$ are mainly determined by the terms
    associated with the differential streaming.
From the identity
   $\rho_\alpha v_{\alpha l} M_p^2/v_{p l}\equiv \rho_p v_{p l} M_\alpha^2/v_{\alpha l}$,
   one can see that for $r \gg r_a$,
   $\xi_\alpha$ and $\xi_p$ tend to have the same magnitude 
   but opposite sign.
At this point, we can also see from the dotted line in Fig.~\ref{fig:angmom}c
    that although the solar rotation introduces
    appreciable difference in the meridional dynamics,
    the Poynting flux never exceeds $0.12\%$ of the total energy budget.
Needless to say, 
    its contribution to the solar wind acceleration
    is in fact determined by its difference between
    1~$R_\odot$ and 1~AU.

At 1~AU, the model yields a total angular momentum loss 
    of
    ${\cal L} =0.17\times 10^{30}$ dyne cm sr$^{-1}$,
    in which the magnetic part
    is ${\cal L}_M= 0.165 \times 10^{30}$ dyne cm sr$^{-1}$,
    consistent with measurements (Pizzo et al.~\cite{Pizzo.etal:83};
    Marsch \& Richter~\cite{MarschRichter:84}).
However, the absolute azimuthal speed $v_{p\phi}$ or $v_{\alpha\phi}$ is larger
    than the measured values
    (although these quantities can only be determined with a modest precision).
Moreover, $\xi_P$ never turns negative, in this sense at variance 
    with the measurements:
    Pizzo et al.~(\cite{Pizzo.etal:83}) and 
    Marsch \& Richter~(\cite{MarschRichter:84})
    showed that
    particles in the fast wind tend to carry a negative angular momentum flux.
Pizzo et al.~(\cite{Pizzo.etal:83}) suggested that the discrepancies between
    the model and measurements may be removed by including
    the stream interaction in the super-Alfv\'enic region.
This is however beyond the scope of this paper.

\section{Concluding remarks}\label{sec:conc}
  
The main aim of this paper is to extend the Weber-Davis analysis (Weber \& Davis~\cite{WD:67})
    on the transport of the angular momentum from the Sun
    by including alpha particles and by allowing the solar wind to flow
    out of the equatorial plane in an axisymmetrical configuration.
Following McKenzie et al.~(\cite{McKenzie.etal:79}),
    we exploit the fact that the gyro-frequency of ions is many orders
    of magnitude higher than any other frequency in ion momentum equations.
From this it follows that the difference between proton and alpha velocities must be
     in the direction of the magnetic field.
Using this alignment condition, the governing equations are then derived from
    the standard five-moment transport equations.

The model equations also enable us to examine quantitatively the effect
    of azimuthal components in limiting the proton-alpha differential streaming in
    the fast wind.
For simplicity, we choose to solve the governing equations on a prescribed poloidal magnetic field line
    located at a colatitude of $70^\circ$ at 1~AU, 
    corresponding to the edge of the fast stream observed by Ulysses
    at solar minimum conditions (McComas~\cite{McComas.etal:00}).
The effects of the azimuthal components on the meridional
    dynamics, if any, are optimal in this regard.
These effects are directly shown by a comparison of two models with and without azimuthal components.

The main results can be summarized as follows:
\begin{enumerate}
\item 
The general analysis concludes that, in agreement with the Weber-Davis model,
    the magnetic field helps the coronal plasma to achieve an effective
    corotation from the coronal base to the Alfv\'enic radius, 
    where the poloidal Alfv\'enic Mach number $M_T=1$ .
$M_T$ has to include the contribution from alpha particles (Eq.~(\ref{eq:mach})).

\item 
In the low latitude fast solar wind,
    the angular momentum loss from the Sun is
    almost entirely due to magnetic stresses.
The proton contribution, which can be as important as the magnetic one in interplanetary space, 
    is offset by alpha particles that develop an azimuthal speed in the direction
    of counter-rotation with the Sun.
\item 
The Poynting flux associated with the azimuthal components is negligible.
Nevertheless, the solar rotation has an appreciable effect
    in limiting the proton-alpha differential streaming in fast solar wind streams
    at low latitudes in interplanetary space.
\end{enumerate}

Although the fast solar wind solution is largely compatible with in situ
    measurements 
    in terms of the ion mass fluxes and terminal speeds,
    it fails in a detailed fashion.
For instance, the model is not able to predict a proton temperature profile
    consistent with UVCS measurements in the inner corona,
    nor does it predict an ion differential speed
    as large as 150~km~s$^{-1}$ at 0.3~AU
    to be comparable with  the Helios observation
    (Marsch et al.~\cite{Marsch.etal:82}).
Hence, including the azimuthal components cannot solely account for the deceleration 
     of alphas relative to protons in interplanetary space.
More sophisticated mechanisms, the ion-cyclotron resonance for instance,
    are expected to alleviate the discrepancies
    (e.g., Li~\cite{Li:03}),
    but can hardly help achieve a satisfactory match
    (e.g., Hu \& Habbal~\cite{HuHabbal:99}).
Nevertheless, such a direction is for sure worth pursuing and is left for a 
    future study.

The model also suffers from the inconsistency that 
    the force balance in the direction perpendicular to
    the poloidal magnetic field 
    is replaced by prescribing a background magnetic field.
In a more rigorous treatment, the poloidal magnetic field
    should be derived self-consistently.
In principle, such a task can be accomplished by
    adopting an iterative approach:
    the parallel and perpendicular force balance are solved
    alternately until a convergence is met (Pneuman \& Kopp~\cite{PK:71};
    Sakurai~\cite{Sakurai:85}).
By doing so, the angular momentum loss from the Sun can be obtained
    self-consistently for all poloidal flux tubes.
An accurate estimate of the duration over which
    the angular momentum of the Sun is completely removed
    is then possible
    (see Hu et al.~\cite{Hu.etal:03b}).

The present paper is aimed at presenting a rather general analysis of the 
    angular momentum loss from a magnetized rotating object
    for flows assuming axial symmetry and incorporating
    two major ion species.
Although for the present Sun, 
    the centrifugal and magnetic forces are so weak that they have
    little impact on the meridional dynamics (especially below the Alfv\'enic point),
    a similar study as presented in the text can be carried out
    for stars that rotate at a faster rate or have a stronger magnetic field
    than the Sun.

\begin{acknowledgements}
This research is supported by a PPARC rolling grant to the University of Wales Aberystwyth.
We thank Shadia Rifai Habbal for her input.
We thank the anonymous referee for his/her comments which helped to improve this paper.
\end{acknowledgements}

\pagebreak

\pagebreak

\pagebreak

\Online
\appendix
\section{Derivation of the governing equations}

In this appendix, it is shown how the 5-moment transport
     equations are reduced to the governing equations
     in section~\ref{sec:goveqs}.
The approach adopted here closely follows that by McKenzie et al.~(\cite{McKenzie.etal:79})
   (see also 
    Hollweg \& Isenberg~\cite{HI:81}).
The original derivation of McKenzie et al.~(\cite{McKenzie.etal:79})
    is restricted to the equatorial flow,
    and ions other than protons are treated as test particles.
Employing the same spirit, we extend their derivation to  
    general flows assuming axial symmetry.
In addition, all ion species are treated on an equal footing, which is 
    particularly important for the solar wind
    since alpha particles can not be seen as test particles.
The central point is that, due to the presence of a strong magnetic field 
    (in the sense that
    the ion gyro-frequency is many orders of magnitude higher than any other frequency
    in the momentum equations),
    the difference vector between proton and alpha velocities must be
    aligned with the magnetic field.

\subsection{General momentum equation}

First of all, let us examine the momentum equation for species ${\sf s}$ 
  (Schunk~\cite{Schunk:77}),
\begin{eqnarray}
&& n_{\sf s} m_{\sf s} \left[
   \frac{\partial \vec{v}_{\sf s}}{\partial t}+ \vec{v}_{\sf s}\cdot\nabla\vec{v}_{\sf s}\right]
     + \nabla p_{\sf s}  \nonumber \\
&& +n_{\sf s} m_{\sf s} \frac{GM_\odot}{r^2}\hat{r}  
     -n_{\sf s} e_{\sf s} \left(\vec{E}+\frac{1}{c}\vec{v}_{\sf s}\times\vec{B}\right)\nonumber \\
&& -\frac{\delta\vec{M}_{\sf s}}{\delta t}
  =0.   \label{eq:vs}
\end{eqnarray}
As usual, species ${\sf s}$ is characterized by its density $n_{\sf s}$,
    velocity $\vec{v}_{\sf s}$,
    mass $m_{\sf s}$, electric charge $e_{\sf s}$ and pressure $p_{\sf s}$.
$e_{\sf s}$ can also be measured in units of electron charge $e$, i.e.,
    $e_{\sf s} = Z_{\sf s} e$ with $Z_e \equiv -1$ by definition.
The momentum exchange rate $\delta \vec{M}_{\sf s}/{\delta t}$ is due to the
    Coulomb frictions.
It is customary to neglect the electron inertia ($m_e = 0$).
As a result,
     the electrostatic field $\vec{E}$ can be expressed as  
\begin{eqnarray}
\vec{E}=-\frac{1}{c}\vec{v}_e\times\vec{B}
   -\frac{\nabla p_e}{n_e e}+\frac{1}{n_e e}\frac{\delta
    \vec{M}_e}{\delta t} . \label{eq:elecfld}
\end{eqnarray}
Substituting the expression for $\vec{E}$ into the magnetic induction law
\begin{eqnarray*}
\frac{\partial \vec{B}}{\partial t} +c\nabla\times\vec{E} =0 ,
\end{eqnarray*}
   one then arrives at 
\begin{eqnarray}
\frac{\partial \vec{B}}{\partial t}-
  \nabla\times\left(\vec{v}_e \times \vec{B}\right) = 0, \label{eq:induct}
\end{eqnarray}
    where $\vec{B}$ is the magnetic field.
The terms in Eq.~(\ref{eq:elecfld}) other than the  motional
    electric field $-\vec{v}_e\times\vec{B}/c$
    are many orders of magnitude smaller and thus have
    been neglected.
All terms have to be kept when $\vec{E}$ is
    substituted into the ion momentum equation however.

The plasma in question consists of two ion species, protons ($p$) and 
    an additional one ($i$).
(Subscript $i$ is used here to indicate that in principle the equations
    to be developed are also applicable if other ion species than alpha particles
    is considered.)
As the frequency in question is well below the electron plasma frequency,
    the expression for $n_e$ follows from quasi-neutrality, 
\begin{eqnarray}
n_e &=& n_p + Z_i n_i .
\end{eqnarray}
Neglecting the displacement current in the Ampere's law, one finds
     the expression for $\vec{v}_e$
\begin{eqnarray}
\vec{v}_e = \frac{n_p\vec{v}_p+ Z_i n_i\vec{v}_i}{n_e}
  -\frac{\vec{j}}{n_e e}, \hspace{0.5cm}
\label{eq:ve}
\end{eqnarray}
where $\vec{j}=c\nabla\times\vec{B}/4\pi$ is the electric current density.

Substitution of Eq.~(\ref{eq:elecfld}) into Eq.~(\ref{eq:vs}) for ion species $k$ ($k=p,i$)
   then leads to
\begin{eqnarray}
&& \frac{\partial \vec{v}_k}{\partial t}+ \vec{v}_k\cdot\nabla\vec{v}_k
     +\frac{\nabla p_k}{n_k m_k} +\frac{Z_k\nabla p_e}{n_e m_k} \nonumber \\
&& +\frac{GM_S}{r^2}\hat{r}  
     -\frac{Z_k}{4\pi n_e m_k}\left(\nabla\times\vec{B}\right)\times\vec{B} \nonumber \\
&& -\frac{1}{n_k m_k}\left[\frac{\delta\vec{M}_k}{\delta t}
     +\frac{Z_k n_k}{n_e}\frac{\delta\vec{M}_e}{\delta t} \right] \nonumber \\
&& +\frac{Z_k e}{m_k c} \frac{n_j Z_j}{n_e} \left(\vec{v}_j-\vec{v}_k\right)\times\vec{B}   
  =0 ,   \label{eq:vk}   
\end{eqnarray}
   where subscript $j$ stands for ion species other than $k$, namely, $j=p$ for $k=i$
   and vice versa.
Note that, when deriving equation~(\ref{eq:vk}), we have used equation~(\ref{eq:ve}) to
   evaluate the electron velocity $\vec{v}_e$ in the expression for $\vec{E}$.
The electric current $\vec{j}$ can be dropped when 
   $\vec{v}_e$ is evaluated elsewhere.
This is because, in the context of the solar wind, $\vec{j}$ is the
   large-scale electric current
   and is negligible since the spatial scale at which the magnetic field evolves 
   is well beyond the proton inertial length.

\subsection{Alignment conditions for electrons}

Now it becomes necessary to separate explicitly the poloidal and azimuthal 
    components of the magnetic field and species velocities, namely,
\begin{eqnarray*}
\vec{B}=\vec{B}_P + B_\phi \hat{\phi}, \hspace{0.2cm}
\vec{v}_{\sf s}=\vec{v}_{{\sf s} P} + v_{{\sf s} \phi} \hat{\phi}, 
\end{eqnarray*}
   where subscript $P$ stands for the poloidal component. 
The assumption of azimuthal symmetry ($\partial/\partial\phi$=0) allows 
  ${\vec B}_P$ to be expressed in terms of the magnetic flux function 
  $\psi(r, \theta; t)$, i.e.,  
\begin{eqnarray*}
\vec{B}_P=\nabla\psi\times\frac{\hat{\phi}}{r\sin\theta}.
\end{eqnarray*}
The magnetic induction law, Eq.~(\ref{eq:induct}), can then be rewritten as 
\begin{eqnarray}
\frac{\partial{\psi}}{\partial t} &+& 
   \vec{v}_{eP}\cdot\nabla\psi=0, \label{eq:psi}\\
\frac{\partial B_\phi}{\partial t} &+&
  r\sin\theta\nabla\cdot\left[\frac{1}{r\sin\theta}
   \left(B_\phi \vec{v}_{eP}-v_{e\phi} \vec{B}_P\right)\right] =0 .  \label{eq:mom_azi}
\end{eqnarray}
For a steady state, from Eq.~(\ref{eq:psi}) follows $\vec{v}_{eP}\cdot\nabla\psi=0$,
    which is equivalent to
\begin{eqnarray}
\vec{v}_{eP}\times\vec{B}_P =0. \label{eq:ploidem}
\end{eqnarray}
In other words, the poloidal components of the electron velocity and magnetic field
   are strictly parallel.
In light of this alignment condition, equation~(\ref{eq:mom_azi}) can be 
   shown to reduce to Eq.~(\ref{eq:bphi}) in the text.

\subsection{Equations cast in the flux tube frame}

It proves useful to work in the flux tube frame, whose base vectors are
\begin{eqnarray*}
\hat{e}_1=\vec{B}_P/B_P, \hat{e}_3=\hat{\phi}, \hat{e}_2=\hat{e}_3\times\hat{e}_1.
\end{eqnarray*}
By definition, the magnetic field has only two components, i.e.,
\begin{eqnarray*}
\vec{B} = B_1 \hat{e}_1 +B_3 \hat{e}_3, 
\end{eqnarray*}

We restrict ourselves to time-independent axisymmetrical flows only.
Taking the dot product of $\hat{e}_1$ with Eq.~(\ref{eq:vk}), one arrives at
\begin{eqnarray}
&& v_{k1}\partial_1 v_{k1} - v_{k3}^2\partial_1\ln r\sin\theta \nonumber \\
&& +{v_{k2}^2} \hat{e}_1\cdot(\hat{e}_2\cdot\nabla\hat{e}_2)
   + v_{k2}\partial_2 v_{k1}
   +v_{k1}v_{k2}\hat{e}_1\cdot(\hat{e}_1\cdot\nabla\hat{e}_2) \nonumber \\
&&   +\frac{1}{n_k m_k} \partial_1 p_k 
   +\frac{Z_k}{n_e m_k}\partial_1 p_e
   +\frac{G M_\odot}{r}\partial_1\ln r \nonumber \\
&&   +\frac{Z_k}{4\pi n_e m_k} B_3
    \left(\partial_1 B_3 +B_3\partial_1 \ln r\sin\theta\right) \nonumber \\
&&  -\frac{n_j}{A_k n_e} c_0 (v_{j1}-v_{k1})
   +\Omega_{k1}\frac{B_3}{B_1}\frac{Z_j n_j}{n_e}(v_{j2}-v_{k2}) =0. \label{eq:vk1}
\end{eqnarray}
Similarly, taking the dot product of Eq.~(\ref{eq:vk}) 
   with $\hat{e}_2$ and $\hat{e}_3$ results in, respectively,
\begin{eqnarray}
&& \frac{v_{k1}^2}{\cal{R}} -v_{k3}^2\partial_2\ln r\sin\theta \nonumber\\
&& + v_{k2}\partial_2 v_{k2} 
   + v_{k1}\partial_1 v_{k2}
   + v_{k1}v_{k2}\hat{e}_2\cdot(\hat{e}_2\cdot\nabla\hat{e}_1)\nonumber\\
&& +\frac{1}{n_k m_k} \partial_2 p_k 
   +\frac{Z_k}{n_e m_k}\partial_2 p_e
   +\frac{G M_\odot}{r} \partial_2\ln r \nonumber \\
&& -\frac{Z_k}{4\pi n_e m_k}
    \left[\frac{B_1^2}{\cal{R}}-
      \left(\partial_2\frac{B_1^2+B_3^2}{2}
         +B_3^2\partial_2\ln r\sin\theta\right)\right] \nonumber \\
&& -\frac{n_j}{A_k n_e} c_0 (v_{j2}-v_{k2}) \nonumber \\
&&   +\Omega_{k1}\frac{Z_j n_j}{n_e}
    \left[(v_{j3}-v_{k3})-(v_{j1}-v_{k1})\frac{B_3}{B_1}\right] =0,\label{eq:vk2}
\end{eqnarray}
and
\begin{eqnarray}
&& v_{k1}\left(\partial_1 v_{k3}+v_{k3}\partial_1\ln r\sin\theta \right) \nonumber\\
&& +v_{k2}\left(\partial_2 v_{k3}+v_{k3}\partial_2\ln r\sin\theta \right) \nonumber\\
&& -\frac{Z_k}{4\pi n_e m_k}
      B_1\left(\partial_1 B_3 + B_3\partial_1\ln r\sin\theta\right)\nonumber \\
&& -\frac{n_j}{A_k n_e} c_0 (v_{j3}-v_{k3}) \nonumber \\
&&   -\Omega_{k1}\frac{Z_j n_j}{n_e}(v_{j2}-v_{k2}) =0, \label{eq:vk3}
\end{eqnarray}
where 
\begin{eqnarray*}
{\cal R}=1 /\hat{e}_2\cdot(\hat{e}_1\cdot\nabla\hat{e}_1) 
\end{eqnarray*}
   is the (signed) curvature radius of the poloidal magnetic field line, while
\begin{eqnarray*}
c_0 = \frac{Z_i^2 n_i}{n_e}\nu_{pe} \Phi_{pe}
   +\frac{n_e}{n_i}\nu_{pi}\Phi_{pi}
   +\frac{A_i n_p}{n_e}\nu_{ie}\Phi_{ie} 
\end{eqnarray*}
   is a coefficient associated with Coulomb frictions.
Here $A_i = m_i/m_p$ is the mass number of species $i$.
In addition, $\Omega_{k1}=Z_k e B_1/m_k c$ is the gyro-frequency for species $k$.
$\partial_n = \hat{e}_n\cdot\nabla$ is the directional derivative operator 
   along $\hat{e}_n$ ($n=1,2,3$).

The ion gyro-frequency is many orders of magnitude higher than 
   any other frequency in the momentum equation.
This has two consequences.
First, from equation~(\ref{eq:vk3}), $v_{j2}-v_{k2}$ is 
   far smaller than $v_{k3}$ from an order-of-magnitude estimate.
Since $v_{e2}=0$ (see Eq.~(\ref{eq:ploidem})),
   both $v_{p2}$ and $v_{i2}$ should be very small and 
   can be safely neglected unless they appear alongside the ion-cyclotron frequency.
Second, Eq.~(\ref{eq:vk2}) leads to
\begin{eqnarray}
v_{i3}-v_{p3}=\frac{B_3}{B_1}\left(v_{i1}-v_{p1}\right) .
\end{eqnarray}
In other words, the ion velocity difference is aligned with the magnetic field.
This is Eq.~(\ref{eq:flowdiff}).

Solving Eq.~(\ref{eq:vk3}) for $v_{j2}-v_{k2}$ and then substituting it into
   Eq.~(\ref{eq:vk1}), one arrives at the poloidal momentum 
   equation (Eq.~(\ref{eq:vks})).
It is interesting to note that the magnetic field does not appear
   explicitly in this equation (except for the term $\tan\Phi$), 
   although it plays an essential 
   role in coupling the azimuthal and meridional motions.
Combining Eq.~(\ref{eq:vk3}) for $p$ and $i$, one can obtain
   Eq.~(\ref{eq:vphi}). 
The partial differentiation with respect to time $t$ in the equation 
   is merely for numerical purpose.

In closing, we note that the $p$ and $i$ versions
    of Eq.~(\ref{eq:vk2}) can be combined to 
    yield a force balance condition in the $\hat{e}_2$ direction,
\begin{eqnarray}
&& \frac{\rho_p v_{p1}^2}{\cal{R}} 
     -\rho_p v_{p3}^2\partial_2\ln r\sin\theta \nonumber\\
&+&\frac{\rho_i v_{i1}^2}{\cal{R}} 
     -\rho_i v_{i3}^2\partial_2\ln r\sin\theta \nonumber\\
&+& \partial_2 p_p + \partial_2 p_i 
   +\partial_2 p_e
   +(\rho_p+\rho_i)\frac{G M_\odot}{r} \partial_2\ln r \nonumber \\
&-& \frac{1}{4\pi}
    \left[\frac{B_1^2}{\cal{R}}-
      \left(\partial_2\frac{B_1^2+B_3^2}{2}
       +B_3^2\partial_2\ln r\sin\theta\right)\right] 
=0. \label{eq:vk2t}
\end{eqnarray}
If further expressing the geometrical coefficient ${\cal R}$,
   the differentiation $\partial_2$ 
   and the Lorentz force
   in terms of the magnetic flux function $\psi$, one
   can eventually derive a second-order quasi-linear partial differential equation (PDE)
   for $\psi$.
This PDE can then be solved by using the approaches proposed
   by Pneuman \& Kopp~(\cite{PK:71})
   or Sakurai~(\cite{Sakurai:85}).

\end{document}